\magnification1200
\baselineskip=1.3\normalbaselineskip

\font\frak=eufm10
\def\g{\hbox{\frak\char'0147}}

\def\s{\hbox{\frak\char'0163}}
\def\p{\hbox{\frak\char'0160}}

\def\l{\hbox{\frak\char'0154}}

\def\h{\hbox{\frak\char'0150}}

\def\t{\hbox{\frak\char'0164}}

\def\q{\hbox{\frak\char'0161}}

\def\1{\hbox{\frak\char'0061}}

\font\block=msbm10
\def\C{\hbox{\block\char'0103}}

\def\R{\hbox{\block\char'0122}}

\def\Z{\hbox{\block\char'0132}}

\def\+{\hbox{\block\char'0156}}
\def\H{\hbox{\block\char'0110}}
\def\-{\hbox{\block\char'0157}}

\font\red=eusm10
\def\F{\hbox{\red\char'0106}}

\def\L{\hbox{\red\char'0114}}
\font\green=msam10
\def\SQ{\hbox{\green\char'0003}}

\font\white=cmbsy10
\def\H1{\hbox{\white\char'0110}}
\def\E1{\hbox{\white\char'0105}}
\def\F1{\hbox{\white\char'0106}}

\font\bigtenrm=cmr10 scaled\magstep1

\font\tenbf=cmbx10 scaled\magstep1
\font\bf=cmbx10 scaled\magstep0

\font\new=eurm10
\def\x{\hbox{\new\char'0170}}
\def\y{\hbox{\new\char'0171}}
\def\p1{\hbox{\new\char'0160}}
\def\h1{\hbox{\new\char'0150}}

\font\frak=eufm10

\def\r{\hbox{\frak\char'0162}}
\def\t{\hbox{\frak\char'0164}}

\font\ninerm=cmr9

\noindent
{\bigtenrm A spinor-like representation of the contact
superconformal algebra  $K'(4)$}

\vskip 0.3in
{\bigtenrm Elena Poletaeva}
\vskip 0.1in
{\it Centre for Mathematical Sciences}

{\it Mathematics, Lund University}

{\it Box 118, S-221 00 Lund, Sweden}

{\it Electronic mail:} elena$@$maths.lth.se

\vskip 0.1in

{\ninerm   
In this work 
we construct an
embedding of a nontrivial central extension 
of the contact superconformal algebra 
${K}'(4)$
into the Lie superalgebra of pseudodifferential symbols on
the supercircle   $S^{1\mid 2}$.
Associated with this embedding is a one-parameter family of
spinor-like tiny irreducible representations of ${K}'(4)$
realized just on 4 fields instead of the usual 16.

\vskip 0.4in
\font\red=eusm10

\noindent{\tenbf I. Introduction}
\vskip 0.2in

Recall that
a {\it superconformal algebra}  is a simple complex
 Lie superalgebra, such that it 
contains the centerless Virasoro algebra (i.e. the Witt algebra)
$Witt = \oplus_{n\in \Z}\C L_n$
 as a subalgebra, and has growth 1.
The $\Z$-graded superconformal algebras are ones for which
$ad L_0$ is diagonalizable with finite-dimensional eigenspaces; see Ref. 1.
In general, 
a  superconformal algebra is a subalgebra of the
Lie superalgebra of all derivations of 
$\C [t, t^{-1}]\otimes \Lambda (N),$
where $\Lambda(N)$ is the Grassmann algebra in $N$ odd variables.

The Lie superalgebra $K(N)$ of contact vector fields with
Laurent polynomials as coefficients is  
characterized by its action on a contact 1-form
(Refs. 1, 2, 3, and 25); it is isomorphic to the $SO(N)$ 
{\it superconformal algebra} (Ref. 4).
$K(N)$ is simple except when $N = 4$. In this case
$K'(4) = [K(4), K(4)]$ is simple.
Note that $K'(N)$ is spanned by $2^N$ fields. It was discovered 
independently in Ref. 3 and Ref. 5 
that the Lie superalgebra of contact vector fields
with polynomial coefficients in 1 even and 6 odd variables
contains an exceptional simple Lie superalgebra 
(see also Ref. 2, Refs. 6, 7, and Refs. 23, 24).
In Ref. 3 the exceptional superconformal algebra $CK_6$ was discovered
as a subalgebra of $K(6)$, and it was 
shown  that the derived Lie superalgebra of divergence-free
derivations of $\C [t, t^{-1}]\otimes \Lambda (2)$,
which is spanned by 8 fields, 
can be realized inside  $K(4)$
using the construction of $CK_6$ inside $K(6)$.

Note that a Lie algebra of contact vector fields
can be realized as
a subalgebra of Poisson algebra; see Ref. 8.
The Poisson algebra of formal Laurent series on
$\dot{T}^*S^1 = T^*S^1\setminus S^1$ has a well-known deformation,
that is the Lie algebra $R$ of pseudodifferential
symbols on the circle. The Poisson algebra can be considered to be the
semiclassical limit of $R$; see Refs. 9, 10, 11, and 12.

In this work we define a family $R_h(N)$ of Lie
superalgebras of pseudodifferential
symbols on the supercircle $S^{1\mid N}$,
where $h\in \rbrack 0, 1]$, which
contracts to the Poisson superalgebra.

For each $h$ we construct
an embedding of a central extension
$\hat{K}'(4)$ into $R_h(2)$. These central extensions are isomorphic 
to one of 3 independent central extensions, 
which are known
for $K'(4)$ (Refs. 1, 2, 13 and 14). The corresponding central
element is $h\in R_h(2)$.
The elements of embeddings of  $\hat{K}'(4)$ are power series in $h$; 
considering their limits as  $h\rightarrow 0$, we obtain an
embedding of $K'(4)$  into the Poisson superalgebra.

The idea of our construction is as follows.
We consider the Schwimmer-Seiberg's
deformation $S(2, \alpha)$
of the  Lie superalgebra of divergence-free 
derivations of $\C [t, t^{-1}]\otimes \Lambda (2)$
(Refs. 15 and 1) 
and observe that the exterior derivations of 
$S'(2, \alpha)$ form an $\s\l(2)$ if $\alpha\in\Z$.
The exterior derivations of $S'(2, \alpha)$  for all
$\alpha\in\Z$ generate a subalgebra of the Poisson superalgebra
isomorphic to the loop algebra
$\tilde {\s\l}(2)$ [${\s\l}(2)$ corresponds to $\alpha = 1$].
We prove that  the family
$S'(2, \alpha)$ for all $\alpha \in \Z$
and $\tilde{\s\l}(2)$ generate a Lie superalgebra isomorphic to $K'(4)$.
The similar construction for each $h\in \rbrack 0, 1]$ gives
an embedding of a nontrivial central extension of ${K}'(4)$:
$$\hat{K}'(4)\subset R_h(2). \eqno (1.1)$$
It is known that the Lie algebra
$R$ has two independent central extensions; see Refs. 9, 10, and 11.
Accordingly,
there exist, up to equivalence, two nontrivial 2-cocycles on
its superanalog $R_{h=1}(N)$.
The 2-cocycle on ${K}'(4)$, which corresponds to the central extension 
$\hat{K}'(4)$
is equivalent to the restriction of
one of the 2-cocycles on $R_{h=1}(2)$.

Finally, the embedding (1.1) for $h = 1$ allows us
to define a new one-parameter family of tiny
irreducible representations of $\hat{K}'(4)$.
Recall that there exists a two-parameter family of representations 
of $K'(N)$ in the superspace spanned by $2^N$ fields.
These representations are defined by the natural action
of $K'(N)$ in the spaces of ``densities''; see Ref. 1.

We obtain representations of $\hat{K}'(4)$, where the value of the central 
charge is equal to 1, 
realized on just 4 fields, instead of the usual 16.

\vskip 0.4in
\noindent{\tenbf II. Superconformal algebras}
\vskip 0.4in

In this section we review the notion of a superconformal algebra
and give the necessary definitions.

A {\it superconformal algebra} is a complex 
Lie superalgebra $\g$ such that
\hfil\break
1) $\g$ is simple,
\hfil\break
2) $\g$ contains the Witt  algebra $Witt = der \C [t, t^{-1}] = 
 \oplus_{n\in \Z}\C L_n$
 with the well-known commutation relations
$$[L_n, L_m] = (n - m)L_{n+m} \eqno(2.1)$$ 
 as a subalgebra,
\hfil\break
3) $ad L_0$ is diagonalizable with finite-dimensional eigenspaces:
$$\g = \oplus_j\g_j, \g_j = 
\lbrace x\in\g \mid [L_0, x] = jx \rbrace,\eqno(2.2)$$
so that $dim \g_j < C$, where $C$ is a constant independent of $j$; see Ref. 1.
The main series of superconformal algebras are $W(N)$ ($N \geq 0$), 
$S'(N, \alpha)$ ($N \geq 2$), and
$K'(N)$ ($N \geq 1$); see  Refs. 1 and 25.
 The corresponding central extensions were classified in Ref. 1;
 see also Refs. 2, 13, 14 and 16.

{\it The superalgebras $W(N)$.} Consider the superalgebra
$\C [t, t^{-1}]\otimes \Lambda (N),$
where $\Lambda(N)$ is the Grassmann algebra in $N$ variables
$\theta_1, \ldots, \theta_N$. Let $p$ be the parity of the homogeneous
element. Let $p(t) = \bar{0}$ and $p(\theta_i) =
\bar{1}$ for $i = 1, \ldots, N$.
By definition
$W(N)$ is the Lie superalgebra of all derivations of
$\C [t, t^{-1}]\otimes \Lambda (N).$
 Let $\partial_i$ stand for $\partial/{\partial\theta_i}$ and 
$\partial_t$ stand for $\partial/{\partial t}$.
Every $D\in W(N)$ is represented by a differential operator,
$$D = f\partial_t + \sum_{i=1}^N f_i \partial_i, \eqno(2.3)$$
where $f, f_i \in \C [t, t^{-1}]\otimes \Lambda (N)$.
$W(N)$ has no nontrivial 2-cocycles
if $N > 2$. If $N = 1$ or $2$, then there exists, up to equivalence,
one nontrivial 2-cocycle on $W(N)$. 

{\it The superalgebras $S(N, \alpha)$.} The Lie superalgebra $W(N)$ contains
 a one-parameter family of Lie
superalgebras $S(N, \alpha)$; see Refs. 15 and 1.
By definition
$$S(N, \alpha) = \lbrace D \in W(N) \mid
Div(t^{\alpha}D) = 0\rbrace \hbox { for } \alpha \in \C.\eqno (2.4)$$
Recall that
$$Div(D) =
\partial_t(f) + \sum_{i=1}^N (-1)^{p(f_i)}
\partial_i (f_i)\eqno (2.5)$$
and
$$Div(fD) = Df + fDiv D, \eqno (2.6)$$
where $f$ is an even function.
Let $S'(N, \alpha) =  [S(N, \alpha), S(N, \alpha)]$ be the derived
superalgebra.
Assume that $N > 1$. If ${\alpha}\not\in\Z$, then
$S(N, \alpha)$ is simple, and if 
${\alpha} \in \Z$, then
$S'(N, \alpha) $ is a  simple ideal of $S(N, \alpha)$ of codimension one
defined from the exact sequence,
$$0\rightarrow S'(N, \alpha)\rightarrow
 S(N, \alpha)\rightarrow \C t^{-\alpha} \theta_1 \cdots  \theta_N
\partial_t\rightarrow 0.\eqno (2.7)$$
Notice that
$$S(N, \alpha)\cong S(N, \alpha + n) \hbox{ for } n\in\Z. \eqno (2.8)$$
There exists, up to equivalence, one nontrivial 2-cocycle on $S'(N, \alpha)$
if and only if $N = 2$; see Ref. 1. Let $\hat{S}'(2, \alpha)$ be the
corresponding central extension of ${S}'(2, \alpha)$. 
Note that ${S}'(2, \alpha)$ is spanned by 4 even fields and 4 odd fields.
Sometimes  the name ``$N = 4$ superconformal algebra'' is used
for $\hat{S}'(2, 0)$; see Refs. 4 and 3.

\vskip 0.1 in

{\it The superalgebras $K(N)$.}
By definition
$$K(N) = \lbrace D \in W(N)\mid D\Omega  = f\Omega \hbox{ for some }
f\in \C[t, t^{-1}]\otimes\Lambda(N)\rbrace, \eqno (2.9)$$
where
$$\Omega = dt - \sum_{i=1}^N \theta_id\theta_i \eqno (2.10)$$
is a contact 1-form; see Refs. 1, 2, 3, and 25.
(See also Ref. 26, where the contact superalgebra $K(m, n)$
 was introduced, and Ref. 24).
 Every differential operator 
$D\in K(N)$ can be represented by a single function,
$$f \in \C[t, t^{-1}]\otimes\Lambda(N): 
f \rightarrow D_f. \eqno(2.11)$$
 Let
$$\Delta(f) = 2f - \sum_{i=1}^N\theta_i\partial_i(f).\eqno(2.12)$$
Then 
$$D_f = \Delta(f)\partial_t + \partial_t(f)\sum_{i=1}^N \theta_i\partial_i
+ (-1)^{p(f)}\sum_{i=1}^N\partial_i(f)\partial_i.\eqno(2.13)$$
Notice that
$$\eqalignno{
&D_{f+g} = D_f + D_g, &(2.14)\cr
&[D_f, D_g] = D_{\lbrace f,g\rbrace},\cr
}$$
where
$$\lbrace f, g\rbrace = \Delta(f)\partial_t(g) - \partial_t(f)\Delta(g)
+ (-1)^{p(f)}\sum_{i=1}^N\partial_i(f)\partial_i(g). \eqno(2.15)$$

The superalgebras $K(N)$ are simple, except when $N = 4$. If $N = 4$,
then the derived superalgebra $K'(4) = [K(4), K(4)]$ is a simple ideal in
$K(4)$ of codimension one defined from the exact sequence
$$0\rightarrow K'(4)\rightarrow
 K(4)\rightarrow \C 
D_{t^{-1} \theta_1\theta_2\theta_3\theta_4}\rightarrow 0. \eqno(2.16)$$
There exists no nontrivial 2-cocycles on $K(N)$ if $N > 4$.
If $N \leq 3$, then there exists, up to equivalence,
one nontrivial 2-cocycle. Let $\hat{K}(N)$ be the corresponding central 
extension of ${K}(N)$. Notice that
$\hat{K}(1)$ is isomorphic to the  Neveu-Schwarz algebra
(Ref. 17),
and $\hat{K}(2) \cong \hat{W}(1)$ is isomorphic to the so-called 
$N = 2$  superconformal algebra; see Ref. 18.
The superalgebra $K'(4)$ has 3 independent central extensions
(Refs. 1, 2, 13 and 14),
which is important for our task.

\vskip 0.4 in
\noindent{\tenbf III. Lie superalgebras of pseudodifferential symbols}
\vskip 0.4 in

Recall that the ring $R$ of pseudodifferential symbols is the ring of the formal series
$$A(t, \xi) = \sum_{-\infty}^na_i(t) {\xi}^i,\eqno (3.1)$$
where $a_i(t)\in \C [t, t^{-1}]$, and the variable $\xi$ corresponds
to $\partial/{\partial t}$; see Refs. 9, 10, 11, and 12.
The multiplication rule in $R$ is determined as follows:
$$A(t, \xi)\circ B(t, \xi) =
\sum_{n\geq 0}{1\over {n!}}\partial^n_{\xi}A(t, \xi)
\partial^n_{t}B(t, \xi). \eqno (3.2)$$
Notice that $R$ is a generalization of the associative algebra of
the regular differential operators on the circle, and the multiplication
rule in $R$, when restricted to the polynomials in $\xi$, coincides with the
multiplication rule for the differential operators.
The Lie algebra structure on $R$ is 
given by
$$[A, B] =  A\circ B - B\circ A, \eqno (3.3)$$
where $A, B\in R$.

The Poisson algebra $P$ of pseudodifferential symbols has the same 
underlying vector space. The multiplication in $P$ is naturally defined.
The Poisson bracket is defined as follows:
$$\lbrace A(t, \xi), B(t, \xi) \rbrace = 
\partial_{\xi}A(t, \xi)\partial_{t}B(t, \xi) -
\partial_tA(t, \xi)\partial_{\xi}B(t, \xi)\eqno (3.4)$$
(Refs. 12 and 19).
One can construct the contraction of the Lie algebra $R$ to $P$ 
using the linear isomorphisms:
$$\varphi_h : R\longrightarrow R\eqno (3.5)$$
defined by
$$\varphi_h(a_i(t) {\xi}^i) = 
a_i(t)h^i{\xi}^i, \hbox{ where } h\in \rbrack 0, 1],\eqno (3.6)$$
see Ref. 12.
The new  multiplication in $R$ is defined by
$$A\circ_h B = \varphi_h^{-1}(\varphi_h (A)\circ\varphi_h (B)).\eqno (3.7)$$
Correspondingly, the commutator is
$$[A, B]_h = A\circ_h B - B\circ_h A. \eqno (3.8)$$
Thus
$$[A, B]_h = h\lbrace A, B \rbrace + hO(h).\eqno (3.9)$$
Hence
$$\hbox{lim}_{h\rightarrow 0} {1\over h}[A, B]_h = \lbrace A, B \rbrace.
\eqno (3.10)$$

To construct a superanalog of $R$, consider an associative superalgebra
$\Theta_h(N)$ with generators
$\theta_1,\ldots, \theta_N, \partial_1,\ldots, \partial_N$
and relations
$$\eqalignno{
&\theta_i\theta_j = - \theta_j\theta_i,\cr
&\partial_i\partial_j = - \partial_j\partial_i, &(3.11)\cr
&\partial_i\theta_j = h\delta_{i,j}  - \theta_j\partial_i,\cr
}$$
where $h\in \rbrack 0, 1]$.
Define an associative superalgebra,
$$R_h(N) = R \otimes \Theta_h(N), \eqno (3.12)$$
such that
$$(A \otimes X) (B \otimes  Y) = {1\over h}(A\circ_h B)\otimes (X Y),
\eqno (3.13)$$
 where $A, B \in R$, and $X, Y \in \Theta_h(N)$.
The product in $R_h(N)$
determines the natural Lie superalgebra structure on 
this space:
$$[(A \otimes X), (B \otimes  Y)]_h = 
{1\over h}(A\circ_h B)\otimes (X Y) 
- (-1)^{p(X)p(Y)}{1\over h} (B\circ_h A)\otimes (Y X). \eqno (3.14)$$
For each $h\in \rbrack 0, 1]$
there exists an embedding
$$W(N)\subset R_h(N), \eqno (3.15)$$
such that the commutation relations in $R_h(N)$,
when restricted to $W(N)$, coincide with the commutation relations in  $W(N)$.
In particular, when $h = 1$, we obtain the superanalog 
$R(N) := R_{h=1}(N)$
of the Lie algebra of pseudodifferential symbols on the circle.

The Poisson superalgebra $P(N)$ has the underlying vector space
$P\otimes \Theta (N)$, where 
\hfil\break
$\Theta (N) := \Theta_{h = 0} (N)$
is the Grassman algebra
with generators
$\theta_1,\ldots, \theta_N, \bar{\theta}_1, \ldots, \bar{\theta}_N$,
where $\bar{\theta}_i = \partial_i$ for $i = 1, \ldots, N$.
The Poisson bracket is defined as follows:
$$\lbrace A, B \rbrace = 
\partial_{\xi}A\partial_{t}B -
\partial_tA\partial_{\xi}B - (-1)^{p(A)}
(\sum_{i=1}^N\partial_{\theta_i}A\partial_ {\bar{\theta}_i}B
+ \partial_ {\bar{\theta}_i}A\partial_{\theta_i}B),\eqno (3.16)$$
where $A, B \in P(N)$; cf. Refs. 2, 5. Thus
$$\hbox{lim}_{h\rightarrow 0} [A, B]_h = \lbrace A, B \rbrace.\eqno (3.17)$$
Correspondingly, we have the embedding
$$W(N)\subset P(N). \eqno (3.18)$$

{\it Remark 3.1:}
Recall that there exist, up to equivalence, two nontrivial 2-cocycles
on $R$ (Refs. 9, 10, and 11).
Analogously, one can define two
2-cocycles, $c_{\xi}$ and $c_t$, on $R(N)$; cf. Ref. 20.
Let $A, B \in R$, and $X, Y \in \Theta_{h=1}(N)$. Then
$$c_{\xi}(A \otimes X, B \otimes  Y) = 
\hbox { the coefficient of } t^{-1}\xi^{-1}\theta_1\ldots\theta_N
\partial_1\ldots\partial_N \eqno (3.19)$$
$$\hbox { in } 
([\log  \xi,A]\circ B)\otimes (X Y),$$
where
$$[\log  \xi,A(t, \xi)] = \Sigma_{k\geq 1}
 {(-1)^{k+1}\over k}\partial^k_tA(t, \xi)\xi^{-k}, \eqno (3.20)$$
and
$$c_t(A \otimes X, B \otimes  Y) = 
\hbox { the coefficient of } t^{-1}\xi^{-1}\theta_1\ldots\theta_N
\partial_1\ldots\partial_N \eqno (3.21)$$
$$\hbox { in } 
([\log  t,A]\circ B)\otimes (X Y),$$
where
$$[\log  t,A(t, \xi)] = \Sigma_{k\geq 1}
 {(-1)^{k+1}\over k}t^{-k}  \partial^k_{\xi}A(t, \xi). \eqno (3.22) $$

\vskip 0.4in
\noindent{\tenbf IV. The construction of embedding}
\vskip 0.4in

Let  $Der S'(2, \alpha)$ be the Lie superalgebra of all derivations of
$S'(2, \alpha)$.
 
{\it Lemma 4.1:} The exterior derivations 
$Der_{ext}S'(2, \alpha)$ for all $\alpha \in\Z$ generate the loop algebra 
$$\tilde{\s\l}(2)\subset P(2).\eqno (4.1)$$

{\it Proof:} In Ref. 21 we observed that 
the exterior derivations of $S'(2, 0)$ form an $\s\l(2)$.
Let $$\lbrace \L_n^{\alpha}, E_n, H_n,
 F_n, \h1_n^{\alpha}, \p1_n^0,
 \x_n^0, \y_n^{\alpha}\rbrace _{n \in\Z}
\eqno (4.2)$$
be  a basis of $S'(2,\alpha)$ defined as follows:
$$\eqalignno{
&\L_n^{\alpha} = -t^n(t \xi + {1\over 2}(n + \alpha + 1)
(\theta_1 \partial_1 +
\theta_2 \partial_2 )), &(4.3)\cr
& E_n = t^n \theta_2 \partial_1, \cr
& H_n = t^n (\theta_2 \partial_2 -
\theta_1 \partial_1), \cr
&F_n = t^n \theta_1 \partial_2, \cr
&\h1_n^{\alpha} = t^{n}\xi\theta_2 -
(n + \alpha)t^{n-1}\theta_1\theta_2\partial_1,\cr
&\p1_n^0 = -t^{n+1}\partial_2,\cr
&\x_n^0 = t^{n+1}\partial_1, \cr
&\y_n^{\alpha} = t^{n}\xi\theta_1 +
(n + \alpha)t^{n-1}\theta_1\theta_2\partial_2.\cr
}$$
Let us show that if  $\alpha \in \Z$, then
$Der_{ext} S'(2,\alpha) \cong 
\s\l(2) = \langle \E1,\H1,\F1 \rangle$, where
$$[\H1, \E1] = 2\E1, 
[\H1, \F1] = - 2\F1,
[\E1, \F1] = \H1, \eqno (4.4)$$
and the action of $\s\l (2)$ is given  as follows:
$$\eqalignno{
&[\E1, \h1_n^{\alpha}] = \x_{n-1+\alpha},
[\E1, \y_n^{\alpha}] = \p1_{n-1+\alpha}^0,
[\F1, \x_n] = \h1_{n+1-\alpha}^{\alpha},
[\F1, \p1_n^0] =  \y_{n+1-\alpha}^{\alpha}, & (4.5)\cr
&[\H1, \x_n^0] = \x_n^0, [\H1, \h1_n^{\alpha}] =
 - \h1_n^{\alpha}, 
[\H1, \p1_n^0] = \p1_n^0, [\H1, \y_n^{\alpha}] =
 - \y_n^{\alpha}.  \cr
}$$
Notice that
$$Der_{ext} S'(2,\alpha) \cong H^1(S'(2,\alpha), S'(2,\alpha)), 
 \eqno(4.6)$$
see Ref. 22. Consider the following $\Z$-grading deg
of $S'(2,\alpha)$:
$$\eqalignno{
&\hbox{deg } \L_n^{\alpha} =  n, \hbox{deg } E_n =  n + 1 - \alpha,
\hbox{deg }  F_n =  n - 1 + \alpha, \hbox{deg } H_n =  n,
&(4.7)\cr
&\hbox{deg } \h1_n^{\alpha} =  n, \hbox{deg } \p1_n =  n,
 \hbox{deg } \x_n =  n + 1 - \alpha, \hbox{deg } \y_n^{\alpha} =  n - 1 + \alpha.
\cr
}$$
Let
$$L_0^{\alpha} =  - \L_{0}^{\alpha} + {1\over 2}(1 -  \alpha)H_{0}.
 \eqno (4.8)$$
Then 
$$[L_0^{\alpha}, s] = (\hbox{deg } s)s \eqno (4.9)$$
for a homogeneous  $s \in S'(2,\alpha)$.
Accordingly,
$$[L_0^{\alpha}, D] = (\hbox{deg } D)D \eqno (4.10)$$
for a homogeneous   $D \in Der_{ext} S'(2,\alpha)$.
On the other hand, since the action of a Lie superalgebra
on its cohomology is trivial, then one must have
$$[L_0^{\alpha}, D] = 0.\eqno (4.11)$$
Hence the nonzero elements of $Der_{ext} S'(2,\alpha)$ have $\hbox{deg} = 0$,
and they preserve the superalgebra
$S'(2,\alpha)_{\hbox{deg}=0}$.
One can   check  that the 
exterior derivations of
$S'(2,\alpha)_{\hbox{deg}=0}$ form an $\s\l(2)$, and  extend them to the
exterior derivations of
$S'(2,\alpha)$  as in (4.5). One should also note that
if the restriction of a derivation of $S'(2,\alpha)$ to $S'(2,\alpha)_{\hbox{deg}=0}$
is zero, then this derivation is inner.
We can identify the exterior derivation
$t^{-\alpha}\xi\theta_1\theta_2$ [see (2.7)] with $-\F1$.
We cannot realize all the exterior derivations as regular 
differential operators on the supercircle,
but can do this using the symbols of
pseudodifferential operators.
In fact,
let $\alpha = 1$. Then
$$Der_{ext}S'(2, 1) = \s\l(2) = \langle \F1, \H1, \E1 \rangle\subset P(2),
\eqno (4.12)$$
where 
$$\F1 = -t^{-1}\xi\theta_1\theta_2,
\H1 =  - \theta_1\partial_1 
- \theta_2\partial_2, 
\E1 =  t\xi^{-1} \partial_1\partial_2. \eqno (4.13) $$
One can then construct the loop algebra of $\s\l(2)$ as follows:
$$\tilde{\s\l}(2) = \langle \F1_n, \H1_n, \E1_n \rangle_{n\in\Z},
 \eqno (4.14)$$
where
$$\eqalignno{
&\F1_n = -t^{n-1}\xi\theta_1\theta_2, &(4.15)\cr
&\H1_n = nt^{n-1}\xi^{-1}\theta_1\theta_2\partial_1\partial_2
-t^{n}(\theta_1\partial_1 + \theta_2\partial_2), \cr
&\E1_n = t^{n+1}\xi^{-1}\partial_1\partial_2.\cr
}$$
The nonvanishing commutation relations are
$$[\H1_n, \E1_k] = 2\E1_{n+k}, 
[\H1_n, \F1_k] = -2\F1_{n+k},  
[\E1_n, \F1_k] = \H1_{n+k}. \eqno (4.16)$$
Let $\alpha \in\Z$. Then 
$$Der_{ext} S'(2, \alpha) \cong 
\langle \F1_{-\alpha+1}, \H1_0, \E1_{\alpha-1}\rangle. \eqno (4.17)$$ 
$$\eqno\SQ$$

{\bf Theorem 4.1:} 
The superalgebras $S'(2, \alpha)$
for all $\alpha\in\Z$ together with $\tilde{\s\l}(2)$ generate a Lie superalgebra 
isomorphic to $K'(4)$.

{\it Proof:} 
Let
$$\eqalignno{
&I_n^0 = t^n(\theta_1\partial_1 + \theta_2\partial_2),\cr
&\r_n = t^{n-1}\theta_1\theta_2\partial_1, &(4.18)\cr
&\s_n = t^{n-1}\theta_1\theta_2\partial_2. \cr
}$$
Then according to (4.3)
$$\eqalignno{
&\L_n^{\alpha} = \L_n^{0} - {1\over 2}\alpha I_n^{0}, &(4.19)\cr
&\h1_n^{\alpha} = \h1_n^{0} - \alpha \r_{n},\cr
&\y_n^{\alpha} = \y_n^{0} + \alpha \s_{n}.\cr
}$$
One can easily check that 
the superalgebras $S'(2, \alpha)$,
where $\alpha\in\Z$, generate $W(2)\subset P(2)$. In fact, 
$W(2)$ is spanned by
8 fields defined in Eq. (4.3), where $\alpha = 0$,  
together with 3 fields defined in Eq. (4.18)
and the field $\F1_n$.
If we include two even fields, $\E1_n$ and $\H1_n$, into the picture,
then from the commutation relations, we obtain two additional odd
fields:
$$\eqalignno{
&\q_n = t^n\xi^{-1}\theta_2\partial_1\partial_2, &(4.20) \cr
&\t_n = -t^n\xi^{-1}\theta_1\partial_1\partial_2.\cr
}$$
Let $\g\subset P(2)$ be the Lie superalgebra generated by the superalgebras
$S'(2, \alpha)$
for all $\alpha\in\Z$ and $\tilde{\s\l}(2)$. 
We will show that there exists an isomorphism:
$$\psi: {K}'(4) \longrightarrow \g. \eqno(4.21)$$
Let
$$\eqalignno{
&\L_n = \L_n^0 + \H1_n + {1\over 2}I_n^0, &(4.22)\cr
&I_n = I_n^0 + \H1_n,\cr
&\p1_n = \p1_n^0 + \t_n,\cr
&\x_n = \x_n^0 - \q_n.\cr
}$$
Set
$$\h1_n = \h1_n^0,\y_n = \y_n^0. \eqno (4.23)$$
Then $\g = \g_{\bar 0}\oplus\g_{\bar 1}$, where
$$\eqalignno{
&\g_{\bar 0} = \langle \L_n, I_n, E_n, H_n, F_n, \E1_n, \H1_n, \F1_n\rangle,
&(4.24)\cr
&\g_{\bar 1} = \langle \h1_n, \p1_n, \x_n, \y_n, 
\r_n, \s_n, \q_n, \t_n\rangle.\cr
}$$
We will describe the nonvanishing commutation relations in $\g$
with respect to this basis.

For $[\g_{\bar 0}, \g_{\bar 0}]$ the relations are:
\hfil\break
$[\L_n, \L_k] = (n - k)\L_{n+k}; 
\qquad\qquad\qquad\qquad\qquad\qquad\qquad\qquad\qquad
\qquad\qquad\qquad\qquad
(4.25)$
\hfil\break
$[H_n,E_k] = 2E_{n+k}, [H_n, F_k] = -2F_{n+k}, [E_n, F_k] = H_{n+k}$;
\hfil\break
$[\H1_n,\E1_k] = 2\E1_{n+k}, [\H1_n, \F1_k] = -2\F1_{n+k}, 
[\E1_n, \F1_k] = \H1_{n+k}$;
\hfil\break
$[\L_n, X_k] = -k X_{n+k}$,
where $X_k = I_k, E_k, H_k, F_k, \E1_k, \H1_k, \F1_k$.

For $[\g_{\bar 0}, \g_{\bar 1}]$ the relations are:
\hfil\break
$[\L_n, X_k] = (-k + {n\over 2})X_{n+k}$,
where $X_k = \h1_k, \p1_k, \x_k, \y_k$;
\qquad\qquad\qquad\qquad 
\qquad\qquad\quad (4.26)
\hfil\break
$[\L_n, X_k] = (-k - {n\over 2})X_{n+k}$,
where $X_k = \r_k, \s_k, \q_k, \t_k$;
\hfil\break
$[I_n, X_k] = nY_{n+k}$, where
$X_k = \h1_k, \p1_k, \x_k, \y_k$, and
$Y_k = \r_k, \t_k, -\q_k, -\s_k$, respectively;
\hfil\break
$[H_n, X_k] =  X_{n+k}$,
where $X_k = \h1_k, \x_k, \q_k, \r_k$;
\hfil\break
$[H_n, X_k] =  - X_{n+k}$,
where $X_k = \y_k, \p1_k, \s_k, \t_k$;
\hfil\break
$[E_n, X_k] =  Y_{n+k}, [F_n, Y_k] =  X_{n+k},$

where $X_k = \y_k, \p1_k, \s_k, \t_k$, and
$Y_k = \h1_k, \x_k, -\r_k, -\q_k$, respectively;
\hfil\break
$[\H1_n, X_k] =  X_{n+k} + nY_{n+k}$,

where  $X_k = \p1_k, \x_k, \q_k, \t_k$, and
$Y_k = \t_k, -\q_k, 0, 0$, respectively;
\hfil\break
$[\H1_n, X_k] =  - X_{n+k} - nY_{n+k}$,

where $X_k =  \h1_k, \y_k, \r_k, \s_k$, and
$Y_k = \r_k, -\s_k, 0, 0$, respectively;
\hfil\break
$[\E1_n, X_k] = Y_{n+k} - nZ_{n+k}$, $[\F1_n, Y_k] = X_{n+k} - n\bar{Z}_{n+k}$,
where
$X_k = \h1_k, \y_k, \r_k, \s_k$,

$Y_k = \x_k, \p1_k, -\q_k, -\t_k$,
$Z_k = \q_k, -\t_k, 0, 0$, and
$\bar{Z}_k = -\r_k, \s_k, 0, 0$, respectively.

Finally, for $[\g_{\bar 1}, \g_{\bar 1}]$ the relations are:
\hfil\break
$[\h1_n, \x_k] =
 (k - n)E_{n+k},
[\p1_n, \y_k] = (k - n)F_{n+k}$,
\qquad\qquad\qquad\qquad
\qquad\qquad\qquad\qquad
 (4.27)
\hfil\break
$[\h1_n, \p1_k] = \L_{n+k}
 - {1\over 2}(k - n)H_{n+k},
 [\x_n, \y_k] = -\L_{n+k} +
 {1\over 2}(k - n)H_{n+k}$,
 \hfil\break
 $[\h1_n, \q_k] = E_{n+k}, [\x_n, \r_k] = E_{n+k}$,
 $[\p1_n, \s_k] = F_{n+k}, [\y_n, \t_k] = F_{n+k}$,
 \hfil\break
 $[\p1_n, \q_k] = -\E1_{n+k}, [\x_n, \t_k] = -\E1_{n+k},
 [\h1_n, \s_k] = -\F1_{n+k},
 [\y_n, \r_k] = -\F1_{n+k}$,
 \hfil\break
$[\p1_n, \r_k] = {1\over 2}I_{n+k} - {1\over 2}(H_{n+k} + {\H1}_{n+k}),
[\x_n, \s_k] = {1\over 2}I_{n+k} + {1\over 2}(H_{n+k} - {\H1}_{n+k})$,
\hfil\break
$[\h1_n, \t_k] = {1\over 2}I_{n+k} + {1\over 2}(H_{n+k} + {\H1}_{n+k}),
[\y_n, \q_k] = {1\over 2}I_{n+k} - {1\over 2}(H_{n+k} - {\H1}_{n+k})$.
\vskip 0.1in

Recall that the elements of  ${K}(4)$ can be identified with the functions
from
\hfil\break
$\C[t, t^{-1}]\otimes\Lambda(4)$.
Let
$$\check\theta_1 = \theta_2\theta_3\theta_4,
\check\theta_2 = \theta_1\theta_3\theta_4,
\check\theta_3 = \theta_1\theta_2\theta_4,
\check\theta_4 = \theta_1\theta_2\theta_3. \eqno(4.28)$$
The following 16 series of functions together with 
$t^{-1}\theta_1\theta_2\theta_3\theta_4$
span $\C[t, t^{-1}]\otimes\Lambda(4)$:
$$\eqalignno{
&f^{1}_n = 2nt^{n-1}\theta_1\theta_2\theta_3\theta_4,&(4.29)\cr
&f^2_n = -{1\over 2}t^{n+1}  + {1\over 2}i
t^n(\theta_2\theta_3  - \theta_1\theta_4)
- {1\over 2}n(n+1) t^{n-1} \theta_1\theta_2\theta_3\theta_4,\cr
&f^k_n = {1\over 2}t^{n\mp 1} (\pm\theta_1\theta_2 \mp \theta_3\theta_4
-i\theta_1\theta_3 - i\theta_2\theta_4), k = 3, 4,\cr
&f^5_n = 
it^{n} (\theta_1\theta_4  - \theta_2\theta_3),\cr
&f^{k}_n =  {1\over 2}t^{n} (\mp\theta_1\theta_4 \mp \theta_2\theta_3
+ i\theta_2\theta_4 - i\theta_1\theta_3), k = 6, 7,\cr
&f^{8}_n = 
-i t^{n} (\theta_1\theta_2  + \theta_3\theta_4),\cr
&f^{k}_n = {(i)^{p(k)}\over {\sqrt 8}}
\bigl(t^{n}(\theta_1 \mp i\theta_2 \mp \theta_3 +i\theta_4)
-nt^{n-1}
(\check\theta_1 \pm i\check\theta_2 \mp \check\theta_3 -i\check\theta_4)\bigr),
 k = 9, 10,   \cr
&f^{k}_n
 = {(i)^{p(k)}\over {\sqrt 8}}\bigl(
t^{n+1}(\theta_1 \pm i\theta_2 \mp \theta_3 - i\theta_4)
-(n+1)t^{n}
(\check\theta_1 \mp i\check\theta_2 \mp \check\theta_3 + i\check\theta_4)\bigr),
k = 11, 12,  \cr
&f^{k}_n = {(-i)^{p(k)}\over {\sqrt 2}}
t^{n-1}
(\check\theta_1 \pm i\check\theta_2 \mp \check\theta_3 - i\check\theta_4),
k = 13, 14,\cr
&f^{k}_n =  {(-i)^{p(k)}\over {\sqrt 2}}
t^{n}
(\check\theta_1 \mp i\check\theta_2 \mp \check\theta_3 + i\check\theta_4),
k = 15, 16,\cr
}$$
where $p(k) = 0$ if $k$ is even, and $p(k) = 1$ if $k$ is odd.

The 16 series of the corresponding differential operators
$\lbrace D_{f^i_n}\rbrace_{i=1,\ldots, 16}$ span ${K}'(4)$.
Set
$$\eqalignno{
&\psi(D_{f^1_n}) = I_n, \psi(D_{f^2_n}) = \L_n,&(4.30)\cr
&\psi(D_{f^3_n}) = E_n, 
\psi(D_{f^4_n}) = F_n,
\psi(D_{f^5_n}) = H_n,\cr
&\psi(D_{f^6_n}) = \E1_n, 
\psi(D_{f^7_n}) = \F1_n,
\psi(D_{f^8_n}) = \H1_n,\cr
&\psi(D_{f^9_n}) =  \x_n,
\psi(D_{f^{10}_n}) = \h1_n,
\psi(D_{f^{11}_n}) = \y_n,
\psi(D_{f^{12}_n}) = \p1_n,\cr
&\psi(D_{f^{13}_n}) = \q_n,
\psi(D_{f^{14}_n}) = \r_n,
\psi(D_{f^{15}_n}) = \s_n,
\psi(D_{f^{16}_n}) = \t_n.\cr
}$$
Notice that $f^{1}_n = 0$, if $n = 0$. This corresponds to the fact 
that $D_{t^{-1}\theta_1\theta_2\theta_3\theta_4} \not\in {K}'(4)$.
One can verify that $\psi$ is an isomorphism from 
${K}'(4)$ onto $\g$.
$$\eqno \SQ$$

{\it Remark 4.2:} We have obtained an embedding 
$$K'(4) \subset P(2).\eqno (4.31)$$
In general, a  Lie algebra of contact vector fields
can be realized as a
subalgebra of Poisson algebra; see Ref. 8. 
We will explain this from the 
geometrical point of view in application to our case.
Recall that the Lie algebra $Vect(S^1)$
of smooth vector fields on the circle has a natural embedding
into the Poisson algebra of functions on 
the cylinder $\dot{T}^*S^1 = T^*S^1\setminus S^1$
with the removed zero section; see Refs. 11, 12 and 19.
One can  introduce 
the Darboux coordinates
$(q, p) = (t, \xi)$ on this manifold.
The symbols of differential operators are functions on  $\dot{T}^*S^1$
which are formal Laurent series in $p$ with coefficients
periodic in $q$. Correspondingly, 
they define Hamiltonian vector fields on $\dot{T}^*S^1$:
$$A(q, p) \longrightarrow H_A = 
\partial_p A\partial_q - \partial_q A\partial_p.\eqno (4.32)$$
The embedding of $Vect(S^1)$ into
the Lie algebra of Hamiltonian vector fields on $\dot{T}^*S^1$ is given by
$$f(q)\partial_q\longrightarrow H_{f(q)p}.\eqno (4.33)$$
Notice that we obtain a subalgebra of
Hamiltonian vector fields  with Hamiltonians 
which are homogeneous of degree 1.
(This condition holds in general, if one considers the
{\it symplectification} of a contact manifold; see Ref. 8.)
In other words, we obtain
a subalgebra of Hamiltonian vector fields,
which commute with the (semi-) Euler vector field:
$$[H_A, p\partial_p] = 0.\eqno (4.34)$$
We will show that for $N\geq 0$
there exists the analogous embedding:
$$K(2N) \subset P(N).\eqno (4.35)$$
The analog of the formula (4.32) in the supercase is
as follows (Refs. 2, 5):
$$A(q, p,{\theta}_i, {\bar{\theta}_i}) \longrightarrow H_A = 
\partial_p A\partial_q - \partial_q A\partial_p
-(-1)^{p(A)}\sum_{i=1}^N(\partial_{{\theta}_i}A\partial_{\bar{\theta}_i} +
\partial_{\bar{\theta}_i} A\partial_{{\theta}_i}).\eqno (4.36)$$
Then $K(2N)$ is defined as the set of all (Hamiltonian) functions
$A(q, p,{\theta}_i, {\bar{\theta}_i})\in P(N)$ such that
$$[H_A, p\partial_{p} 
+ \sum_{i=1}^N\bar{\theta}_i\partial_{\bar{\theta}_i}] = 0.\eqno (4.37)$$
Equivalently, we have the following
characterization of the embedding (4.35).
Consider a $\Z$-grading of the (associative) superalgebra 
$P(N) = \oplus_{j\in\Z}P_j(N)$ defined by
$$\eqalignno{
&\hbox{deg } p = \hbox{deg } \bar{\theta}_i = 1 \hbox{ for }i = 1, \ldots, N,  &(4.38)\cr
&\hbox{deg } q = \hbox{deg } {\theta}_i = 0 \hbox{ for }i = 1, \ldots, N.\cr
}$$
Thus with respect to the Poisson bracket,
$$\lbrace P_j(N), P_k(N)\rbrace \subset P_{j+k-1}(N).\eqno (4.39)$$
Then 
$$K(2N) = P_1(N).\eqno (4.40)$$

{\bf Theorem 4.2:}
There exists an embedding,
$$\hat{K}'(4) \subset R_h(2),\eqno (4.41)$$
for each $h\in \rbrack 0, 1]$, such that
the central element in $\hat{K}'(4)$ is $h\in R_h(2)$, and
$$\hbox{lim}_{h\rightarrow 0}\hat{K}'(4) =  K'(4)\subset P(2).\eqno (4.42)$$

{\it Proof:} For each $h\in \rbrack 0, 1]$ and $\alpha \in\Z$
we have an embedding,
$$DerS'(2, \alpha)\subset R_h(2).\eqno (4.43)$$
The exterior derivations 
$Der_{ext}S'(2, \alpha)$ for all $\alpha \in\Z$ generate the loop algebra, 
$$\tilde{\s\l}(2) =\langle \F1_n, \H1_n, \E1_n \rangle_{n\in\Z}
\subset R_h(2), \hbox{ where }\eqno (4.44)$$
$$\eqalignno{
&\F1_n = -t^{n-1}\xi\theta_1\theta_2, &(4.45)\cr
&\H1_n = {1\over h} ((\xi^{-1}\circ_h t^n\xi)
(h^2 - h\theta_1\partial_1 - h\theta_2\partial_2 -
\theta_1\theta_2\partial_1\partial_2) + 
t^n\theta_1\theta_2\partial_1\partial_2), \cr
&\E1_n = (\xi^{-1}\circ_h t^{n+1})\partial_1\partial_2,\cr
}$$
so that Eqs. (4.16)-(4.17) hold.
Let $\g\subset R_h(2)$ be the Lie superalgebra generated by 
$S'(2, \alpha)$ for all $\alpha\in\Z$ 
and $\tilde{\s\l}(2)$.
Set
$$\eqalignno{
&\q_n = (\xi^{-1}\circ_h t^n)
(h\partial_1 + \theta_2\partial_1\partial_2),  &(4.46)\cr
&\t_n = (\xi^{-1}\circ_h t^n)
(h\partial_2 - \theta_1\partial_1\partial_2).\cr
}$$
The basis (4.24)  in $\g$ is defined by 
Eqs. (4.3), (4.18), (4.22)-(4.23) and (4.45)-(4.46).
The commutation relations in $\g$ with respect to this basis
are given by Eqs.
(4.25)-(4.27). The Lie superalgebra $\g$ is isomorphic to a 
central extension,
$$\hat{K}'(4) = {K}'(4) \oplus \C C\eqno(4.47)$$
of ${K}'(4)$. The corresponding
2-cocycle (up to equivalence) is 
$$\eqalignno{
&c(t^{n+1}, t^{k+1}\theta_1\theta_2\theta_3\theta_4) = \delta_{n+k+2,0},
&(4.48)\cr
&c(t^{n+1}\theta_i, t^{k+1}\partial_i(\theta_1\theta_2\theta_3\theta_4))
 = {1\over 2}\delta_{n+k+2,0} \hbox{ for } i = 1, \ldots, 4.\cr
}$$
The isomorphism,
$$\psi: \hat{K}'(4) \longrightarrow \g \eqno(4.49)$$
is defined by Eq. (4.30) and the equation
$$\psi(C) = I_{0} = h\in R_h(2). \eqno(4.50)$$
The corresponding 2-cocycle in the basis (4.24) is
$$\eqalignno{
&c(\p1_n, \r_k) = {1\over 2}\delta_{n, -k}, &(4.51)\cr
&c(\x_n, \s_k) = {1\over 2}\delta_{n, -k},\cr
&c(\h1_n, \t_k) = {1\over 2}\delta_{n, -k},\cr
&c(\y_n, \q_k) = {1\over 2}\delta_{n, -k},\cr
&c(\L_n, I_k) = n\delta_{n, -k}.\cr
}$$
Note that in the realization of ${K}'(4)$ inside $P(2)$,
obtained in Theorem 4.1, we have $I_{0} = 0$.
$$\eqno \SQ$$

{\it Remark 4.3:} The 2-cocycle $c$  is one of three
nontrivial 2-cocycles on ${K}'(4)$; see Refs. 1 and 2.  
[In Ref. 1 this cocycle is defined by  Eq. (4.22), where
$d = 0, e = 1$].
Note that the cocycle  $c$ is equivalent to the restriction of the 
2-cocycle $c_{t}$ on $R(2)$; see Eqs. (3.21), (3.22).

\vskip 0.4 in
\noindent{\tenbf V. One-parameter family 
of representations of $\hat{K}'(4)$}
\vskip 0.4 in

{\bf Theorem 5.1:}
There exists a one-parameter family of irreducible
representations of $\hat{K}'(4)$
depending on parameter $\mu \in \C$
in the superspace spanned by 2 even fields and 2 odd fields
where the value of the central charge is equal to one.

{\it Proof:} 
Let $g\in t^{\mu}\C [t, t^{-1}]$, where 
$\mu \in \R \setminus\Z$.
One can think of $\xi^{-1}$ as 
the anti-derivative,  
$$\xi^{-1}g(t) = \int g(t)dt. \eqno (5.1)$$ 
Let $f(t) \in \C [t, t^{-1}]$.
According to (3.2),
$$\xi^{-1}\circ f= \sum_{n=0}^{\infty}(-1)^n(\xi^{n}f)\xi^{-n-1}.\eqno (5.2)$$
Notice that this formula, when applied to a function $g$,
corresponds to the formula of integration by parts.
Let
$$V^{\mu} = t^\mu\C [t, t^{-1}]\otimes \Lambda (2) = 
t^\mu\C [t, t^{-1}]\otimes \langle 1, \theta_1, \theta_2, 
\theta_1\theta_2 \rangle, \hbox{ }\mu \in \R \setminus\Z.\eqno (5.3)$$
Using the realization of $\hat{K}'(4)$ inside $R(2)$  
(see Theorem 4.2 for $h = 1$) we obtain a representation of $\hat{K}'(4)$
in  $V^{\mu}$. A central element in $\hat{K}'(4)$
is $I_0 = 1\in R(2)$; the 2-cocycle is defined by Eq. (4.51).
Let $\lbrace v_m^i\rbrace$, where ${m\in \Z}$ and $i = 0, 1, 2, 3$,
be the following basis in $V^{\mu}$:
$$\eqalignno{
&v_m^0 = {1\over {m + \mu}}t^{m + \mu}, &(5.4)\cr
&v_m^1 = t^{m + \mu}\theta_1, \cr
&v_m^2 = t^{m + \mu}\theta_2, \cr
&v_m^{3} = t^{m + \mu}\theta_1\theta_2. \cr
}$$
The action of $\hat{K}'(4)$ is given as follows:
$$\eqalignno{
&\L_n(v_m^0) = -(m + n + \mu - 1)v_{m+n}^0,&(5.5)\cr
&\L_n(v_m^i) = -(m + {1\over 2}n + \mu)v_{m+n}^i, i = 1, 2,\cr
&\L_n(v_m^{3}) = -(m + n + \mu + 1)v_{m+n}^{3},\cr
&E_n(v_m^{1}) = v_{m+n}^{2}, F_n(v_m^{2}) = v_{m+n}^{1},\cr
&\E1_n(v_m^{3}) =  v_{m+n+2}^{0}, \F1_n(v_m^{0}) =  - v_{m+n-2}^{3},\cr
&H_n(v_m^{i}) = \mp v_{m+n}^{i}, i = 1, 2,\cr
&\H1_n(v_m^{i}) = \pm v_{m+n}^{i}, i = 0, 3,\cr
&\h1_n(v_m^{1}) =  - (m + n + \mu) v_{m+n-1}^{3},
\y_n(v_m^{2}) =  (m + n + \mu) v_{m+n-1}^{3},\cr
&\h1_n(v_m^{0}) = v_{m+n-1}^{2},
\y_n(v_m^{0}) = v_{m+n-1}^{1},\cr
&\x_n(v_m^{1}) =  (m + n + \mu)v_{m+n+1}^{0},
\p1_n(v_m^{2}) = - (m + n + \mu)v_{m+n+1}^{0},\cr
&\x_n(v_m^{3}) = v_{m+n+1}^{2},
\p1_n(v_m^{3}) = v_{m+n+1}^{1},\cr
&\r_n(v_m^{1}) = v_{m+n-1}^{3},
\s_n(v_m^{2}) = v_{m+n-1}^{3},\cr
&\q_n(v_m^{1}) = v_{m+n+1}^{0},
\t_n(v_m^{2}) = v_{m+n+1}^{0},\cr
&I_n(v_m^{i}) = v_{m+n}^{i}, i = 0, 1, 2, 3.\cr
}$$
Note that $I_0$ acts by the identity operator.
One can then define a one-parameter family of representations of $\hat{K}'(4)$
depending on parameter $\mu \in \C$
in the superspace $V = \langle v_m^{0}, v_m^{3}, v_m^{1}, v_m^{2}
\rangle_{m \in \Z}$,
where $p(v_m^{i})  = \bar{0}$, for $i = 0, 3$, and
$p(v_m^{i}) = \bar{1}$ for $i = 1, 2$,
according to the formulas (5.5). 
$$\eqno\SQ$$

{\it Remark 5.1:}
The elements $\lbrace\L_n, H_n, \h1_n, \p1_n\rbrace_{n\in\Z}$
span a subalgebra of 
${K}'(4)$ isomorphic to  $K(2)$.
Note that
$V$ decomposes into the direct sum of two
submodules over this superalgebra:
$$V = \langle v_m^{0}, v_m^{2}\rangle_{m\in\Z} \oplus
\langle v_m^{3}, v_m^{1}\rangle_{m\in\Z}.\eqno (5.6)$$

{\it Remark 5.2:}
We conjecture that there exists 
 a {\it two}-parameter family of representations of
$\hat{K}'(4)$ in the superspace spanned by 4 fields. 
In order to define it, instead of the superspace of functions, $V^{\mu}$,
one should  consider the superspace of ``densities''.

\vskip 0.2in
\noindent   
{\tenbf Acknowledgments} 

Part of this work was done while I was visiting
the Max-Planck-Institut f\"ur  Mathematik  in Bonn (Ref. 27).
I wish to thank MPI for the hospitality and  support.
I am grateful to  B. Feigin, A. Givental, I. Kantor, B. Khesin,
D. Leites, C. Roger, V. Serganova,  and I. Zakharevich
for very useful discussions.

When the paper was in print I found out that
Refs. 23, 24, 25, and 26 were missing.
I would like to thank V.G. Kac for reading my work and clarifying
that some corrections have to be made in regard to the references. 

\vskip 0.2in
\font\red=cmbsy10
\def\~{\hbox{\red\char'0016}}

\item{[1]}
V. G. Kac and J. W. van de Leur,
``On classification of superconformal algebras'', in
{\it Strings-88}, edited by S. J. Gates {\it et al}.
(World Scientific, Singapore, 1989), pp. 77-106.

\item{[2]} P. Grozman, D. Leites, and I. Shchepochkina,
``Lie superalgebras of string theories'',

hep-th/9702120.

\item{[3]}
S.-J. Cheng and V. G.  Kac,
``A new $N = 6$ superconformal algebra'',
Commun. Math. Phys. {\bf 186},  219-231 (1997).

\item{[4]} M. Ademollo, L. Brink, A. D'Adda {\it et al}.,
``Dual strings with $U(1)$ colour symmetry'', 
Nucl. Phys. B {\bf 111}, 77-110 (1976).

\item{[5]} I. Shchepochkina,
``The five exceptional simple Lie superalgebras of vector fields'',

hep-th/9702121.

\item{[6]} I. Shchepochkina,
``The five exceptional simple Lie superalgebras of vector fields'',
Funkt. Anal. i Prilozhen. {\bf 33},  59-72 (1999). 
[Funct. Anal. Appl. {\bf 33},  208-219 (1999)].

\item{[7]} I. Shchepochkina,
``The five exceptional simple Lie superalgebras of vector fields
and their fourteen regradings'',
Represent. Theory {\bf 3}, 373-415 (1999).

\item{[8]} V. I. Arnold, 
{\it{Mathematical Methods of Classical Mechanics}}
(Springer-Verlag, New York, 1989).

\item{[9]}
O. S. Kravchenko and B. A. Khesin,
``Central extension of the algebra of 
pseudodifferential symbols'',
Funct. Anal. Appl. {\bf 25},  83-85 (1991).

\item{[10]}
B. Khesin and I. Zakharevich,
``Poisson-Lie group of pseudodifferential symbols'',
Commun. Math. Phys. {\bf 171},  475-530 (1995).

\item{[11]}
B. Khesin, V. Lyubashenko, and C. Roger,
``Extensions and contractions of the Lie algebra of
q-pseudodifferential symbols on the circle'',
J. Funct. Anal. {\bf 143},  55-97 (1997).

\item{[12]}
V. Ovsienko and C. Roger,
``Deforming the Lie algebra of vector fields on $S^1$
inside the Lie algebra of pseudodifferential symbols on $S^1$'',
Am. Math. Soc. Trans. {\bf 194}, 211-226 (1999).

\item{[13]} K. Schoutens,
``A non-linear representation of the $d = 2$ $so(4)$-extended
superconformal algebra'',
Phys. Lett. B {\bf 194}, 75-80 (1987).

\item{[14]} K. Schoutens,
``$O(N)$-extended
superconformal field theory in superspace'',
Nucl. Phys. B {\bf 295}, 634-652 (1988).

\item{[15]} A. Schwimmer and N. Seiberg,
``Comments on the $N = 2, 3, 4$ superconformal algebras in two
dimensions'',
Phys. Lett. B {\bf 184}, 191-196 (1987).

\item{[16]} B. Feigin and D. Leites,
``New Lie superalgebras of string theories'',
in {\it Group-Theoretical Methods in
Physics}, edited by  M. Markov {\it et al.},
(Nauka, Moscow, 1983), Vol. 1, pp. 269-273.
[English translation  Gordon and Breach, New York, 1984).

\item{[17]}
A. Neveu and J. H. Schwarz,
``Factorizable dual models of pions'',
Nucl. Phys. B {\bf 31}, 86-112 (1971).

\item{[18]} B. L. Feigin, A. M.  Semikhatov, and I. Yu. Tipunin,
``Equivalence between chain categories of representations of affine
$\s\l (2)$ and $N = 2$ superconformal algebras'', 
J. Math. Phys. {\bf 39},  3865-3905 (1998).

\item{[19]}
V. Ovsienko and C. Roger,
``Deforming the Lie algebra of vector fields on $S^1$
inside the Poisson algebra on $\dot{T}^*S^1$'',
Commun. Math. Phys. {\bf 198}, 97-110 (1998).

\item{[20]} A. O. Radul,
``Non-trivial central extensions of Lie algebra of differential
operators in two and higher dimensions'',
Phys. Lett. B {\bf 265}, 86-91 (1991).

\item{[21]} E. Poletaeva, 
``Semi-infinite cohomology and superconformal algebras'',
C. R. Acad.  Sci., Ser. I: Math {\bf 326}, 533-538 (1998).

\item{[22]} D. Fuks,  {\it{Cohomology of Infinite-Dimensional
Lie Algebras}} (Consultants Bureau, New York,
1986).

\item{[23]}
V. G. Kac,
``Classification of infinite-dimensional simple linearly
compact Lie superalgebras'',
Adv. Math. {\bf 139}, 1-55 (1998).

\item{[24]}
V. G. Kac,
``Structure of some $\Z$-graded Lie superalgebras of vector fields'',
Transform. Groups {\bf 4}, 219-272 (1999).

\item{[25]}
V. G. Kac,
``Superconformal algebras and transitive group actions on quadrics'',
Commun. Math. Phys. {\bf 186}, 233-252 (1997).

\item{[26]}
V. G. Kac,
``Lie superalgebras'',
Adv. Math. {\bf 26}, 8-96 (1977).

\item{[27]} E. Poletaeva, 
``Superconformal algebras and Lie superalgebras of 
the Hodge theory'',

preprint MPI 99-136.

\bye